\documentclass[9pt,twocolumn,twoside]{opti_preprint}
\setboolean{shortarticle}{true}

\usepackage{color}
\usepackage{amsmath}

\title{Overcoming the acoustic diffraction limit in photoacoustic imaging by localization of flowing absorbers}

\author[1,2]{Sergey Vilov}
\author[1,2]{Bastien Arnal}
\author[1,2,*]{Emmanuel Bossy}

\affil[1]{Univ. Grenoble Alpes, LIPHY, F-38000 Grenoble, France}
\affil[2]{CNRS, LIPHY, F-38000 Grenoble, France}

\affil[*]{Corresponding author: emmanuel.bossy@univ-grenoble-alpes.fr}

\begin{abstract}
The resolution of photoacoustic imaging deep inside scattering media is limited by the acoustic diffraction limit. In this work, taking inspiration from super-resolution imaging techniques developed to beat the optical diffraction limit, we demonstrate that the localization of individual optical absorbers can provide super-resolution photoacoustic imaging well beyond the acoustic diffraction limit. As a proof-of-principle experiment, photoacoustic cross-sectional images of microfluidic channels were obtained with a 15 MHz linear CMUT array while absorbing beads were flown through the channels. The localization of individual absorbers allowed to obtain super-resolved cross-sectional image of the channels, by reconstructing both the channel width and position with an accuracy better than $\lambda/10$. Given the discrete nature of endogenous absorbers such as red blood cells, or that of exogenous particular contrast agents, localization is a promising approach to push the current resolution limits of photoacoustic imaging.
\end{abstract}

\begin{document}
\maketitle
\ifthenelse{\boolean{shortarticle}}{\abscontent}{}


Photoacoustic imaging is a multi-wave biomedical imaging modality, based on the detection of ultrasound following light absorption, which therefore provides optical images  with specific absorption contrast~\citep{beard2011biomedical,wang2012photoacoustic}. The resolution of photoacoustic imaging is limited either by optical diffraction or by acoustic diffraction. The optical-resolution regime is limited by optical scattering to the depth range of optical microscopy based on ballistic photons, i.e to depths less than a few hundreds of microns. At larger depths, in the regime of multiply scattered light, the resolution of photoacoustic imaging is limited by acoustic diffraction. Because ultrasound attenuation increases with frequency, the acoustic resolution decreases with depth, and it is widely considered that the depth-to-resolution ratio is on the order of 200 for depth ranging from a few hundreds of micron to several centimeters. Therefore, exactly as for pulse-echo ultrasound imaging, acoustic-resolution photoacoustic imaging is limited at a given depth by the acoustic-diffraction limit that corresponds to the highest detectable frequency.

In recent years, several research groups investigated new approaches to overcome the acoustic-diffraction limit, both for ultrasound imaging and photoacoustic imaging. In pulse-echo ultrasound imaging, many studies took inspiration from localization approaches developped in optics (such as photoactivated localization microscopy (PALM~\cite{betzig2006imaging}) or stochastic optical reconstruction microscopy (STORM~\cite{rust2006sub})). Localization-based imaging techniques are relying on the possibility with a diffraction-limited imaging system to determine the position of a point source with a precision much larger than the size of the point spread function (PSF), 
provided that the PSFs corresponding to different sources are separated in some parameter space~\citep{betzig1995proposed}.  The first proof-of-concept experiments in ultrasound imaging performed localization by detecting the backscattered signals from a diluted solution of microbubbles, and images of tube-based phantoms were reconstructed by the localization of flowing microbubbles through the tubes~\cite{viessmann2013acoustic,oreilly2013super}. Shortly afterwards, Desailly and colleagues proposed to use sono-activated ultrasound contrast agents to perfom ultrasound localization microscopy~\cite{desailly2013sono}: this approach did not require the use of diluted solutions of contrasts agents to fulfill the localization condition, as the approach relied on the localization of randomly and sparsely activated contrast agents. This pioneer work triggered many subsequent works~\cite{errico2015ultrafast,christensen2015vivo,luke2016super,bar2017fast}, including the \textit{in vivo} demonstration by the same group of ultrasound localization microscopy for deep super-resolution vascular imaging in rodents brains~\cite{errico2015ultrafast}. In the photoacoustics community, imaging beyond the acoustic diffraction limit was first investigated with wavefront shaping approaches~\cite{conkey2015super}, and was  further investigated with a method inspired by the SOFI approach in optics~\citep{chaigne2016super}: in SOFI, it was demonstrated by Dertinger and colleagues that a high-order statistical analysis of fluctuating optical images could overcome the diffraction limit, under the assumption that fluctuations were produced by statistically uncorrelated blinking from fluorescent probes~\citep{dertinger2009fast}. This principle was first implemented in photoacoustic imaging with fluctuations produced via multiple speckle illumination~\citep{chaigne2016super}, and more recently with flow-induced fluctuations~\citep{chaigne2017super}. The principles of SOFI have also been applied to pulse-echo ultrasound with fluctuations induced by flowing microbubbles~\citep{bar2017fast}. Super-resolution based on sparsity constraints, as first demonstrated in optics~\cite{gazit2009super}, has also been recently applied to photoacoustic imaging~\cite{murray2017super,hojman2017photoacoustic}.

Overcoming the acoustic diffraction limit in photoacoustic imaging with a localization approach was proposed for the first time to our knowledge by Iskander-Rizk and colleagues~\citep{iskander2017photoacoustic}, who used the difference between absorption spectra of two unresolved absorbers to separate and then localize the absorbers. However, their demonstration was limited to image only two spectrally distinct absorbers within each PSF. 
In this work, we transpose the approach with flowing particles proposed in~\citep{viessmann2013acoustic} for ultrasound imaging to photoacoustic imaging: we demonstrate experimentally that the localization of optical absorbers flowing through microfluidic-based vessel phantoms allows the reconstruction of the sample structure beyond the acoustic diffraction limit.

\begin{figure}[htbp]
\centering
\fbox{\includegraphics[width=\linewidth]{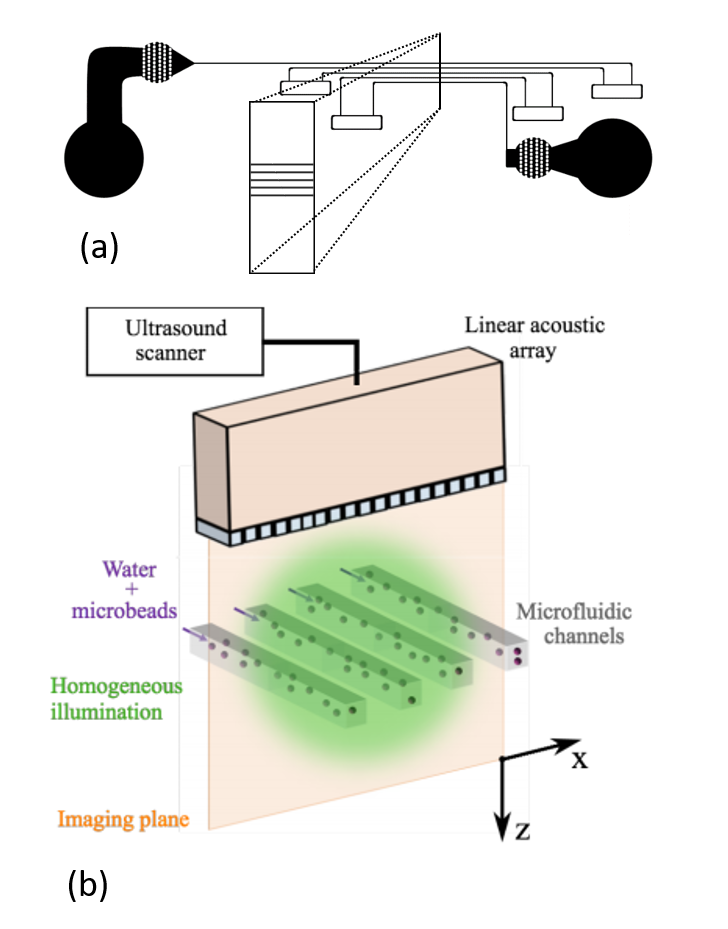}}
\caption{Experimental setup. (a) Overview of the microfluidic circuit, showing the various constitutive elements of the circuit (including input and ouput ports, dust filters, mixers). . The sample to be imaged consists of 5 parallel portions (see zoom) of the circuit. Neighbooring channels in the imaged area are 180 $\mu m$ apart center-to-center, and each channel has a width of 40-$\mu m$ (x direction) and a height of 50-$\mu$m (z direction). (b) Photoacoustic signals were measured with 128 elements of a linear CMUT array (15 MHz, element pitch 100 $\mu m$), originating from 10-$\mu m$ diameter absorbing beads flowing through the channels.} 
\label{fig:exp_setup}
\end{figure}

A schematic of the experimental setup used is shown in Figure 1. The sample to image was made of a microfluidic circuit built in PDMS (polydimethylsiloxane) with standard soft-lithography manufacturing technology~\cite{tang2010basic_bis}. The objective of the experiment was to build super-resolved 2D photoacoustic images of a cross-section of the microfludic circuit. The circuit geometry, illustrated on Fig. 1(a), is designed such that 5 parallel portions of the circuit cross the imaging plane.  In the plane of interest, the dimensions of the channels are 50 $\mu$m in height and 40 $\mu$m in width, and the center-to-center distance between adjacent channels is 180 $\mu$m. As detailed further, the image reconstruction is based on the photoacoustic localization of 10-$\mu$m diameter absorbing microbeads (Microparticles GmbH, Berlin, Germany) flowing through the circuit. The microbeads were circulated through the circuit with a syringe pump (KDS Legato 100, KD Scientific, Holliston,MA, USA) to control the flow. Two-dimensional photoacoustic images of flowing absorbing microbeads were obtained with a linear Capacitive Micromachined Ultrasonic Transducer (CMUT) array (L22-8v, Verasonics, Kirkland,Washington, USA) connected to a multi-channel acquisition electronics (High Frequency Vantage 256, Verasonics, Kirkland,Washington, USA). The sample was illuminated with 5 ns laser pulses at a 100 Hz repetition rate ($\lambda$ = 532 nm, fluence = 3.0 mJ/cm$^2$) from a frequency-doubled Nd:YAG laser (Spitlight DPSS 250, Innolas Laser GmbH,Krailling, Germany). For each laser shot, photoacoustic signals with a center frequency around 15 MHz were acquired simultaneously on 128 elements of the array (pitch 100 um, elevation focus 15 mm). The concentration of microbeads in the circuit was set such that statistically there was only one bead in the field of view at each laser shot.  The PSF of the imaging system, estimated experimentally from the measurement of a single isolated microbead, is shown on Figure 2. The lateral dimension of the PSF, defined as the lateral FWHM of the central lobe, was  178 $\mu$m. The axial dimension of the PSF, defined as the FWHM of the PSF envelope along the axial dimension, was  137 $\mu$m. The photoacoustic images were reconstructed with a conventional delay-and-sum algorithm, on a reconstruction grid with a step size of 5 $\mu$m. As illustrated on Fig. 3(b), showing the conventional photoacoustic image of the 5 channels, the resolution was too poor to resolve individual channels.

\begin{figure}[h!]
\centering
\fbox{\includegraphics[width=\linewidth]{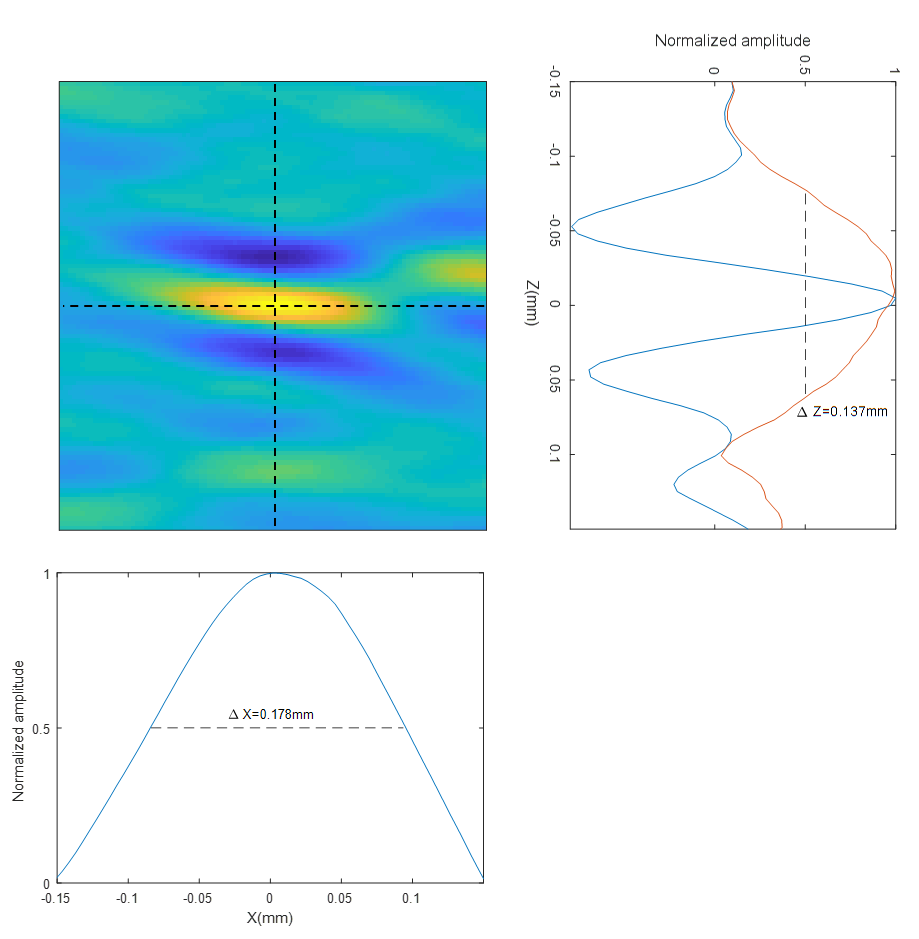}}
\caption{Experimental point spread function (PSF) of the imaging setup, estimated from the signals of a single 10-$\mu$m diameter bead located in one of the microfluidic channels.  The lateral dimension of the PSF, defined as the lateral FWHM of the central lobe, was approximately 178 $\mu$m. The axial dimension of the PSF, defined as the FWHM of the PSF envelope along the axial dimension, was approximately 137 $\mu$m}. 
\label{fig:false-color}
\end{figure}

For the reconstruction of the localization  image, the following procedure was followed: under the assumption that there was a maximum of one bead per PSF on each photoacoustic image (i.e. there was never two beads simultaneoulsy on two adjacent channels), each absorbing bead was localized by detecting local maxima of the 2D cross-correlations between photoacoustic image obtained at each shot laser and the PSF.  The localization of maxima on cross-correlation images rather than on photoacoustic images was used as a spatial matched-filtering approach to improve the localization precision. The localization image was then built as a probability map by computing a 2D histogram of the positions obtained by the localization step. The histogram was built on a grid with a bin size of 1 $\mu$m x 1 $\mu$m, and further smoothed by averaging over a 25 $\mu$m x 25 $\mu$m kernel. The corresponding localization image is shown in Fig. 3(c), with the five channels clearly separated. The average measured center-to-center distance between channels estimated from the peaks centers was 178 $\mu$m, in excellent agreement with the value of 180 $\mu$m expected from the manufacturing process. In addition to discriminate neighbouring channels otherwise blurred by the acoustic diffration limit, the localization of flowing particles into channels may also provide information on the flow itself. Because our setup was limited to cross-sectional imaging, it was not possible at that stage to measure directly the distribution of flow velocity along the channels. Nevertheless, the distribution of microbeads within a channel cross-section is dictated by the channel width and the beads diameter, the water flow profile and the hydrodynamic interactions of the microbeads with the flow and channel boundaries. Some accurate modeling would be needed to derive the flow profile from the positions of the bead, which was out of the scope of this work. Estimating the channel width from the FWHM on the localization images provided values ranging from 40 to 45 $\mu m$, very close to the expected value of 40 $\mu m$. This very good quantitative agreement may however result from a compensation of two effects: the use of a filtering kernel (to smooth spatial variation) leads to overestimate the distribution width, while the finite dimension of the microbeads diameter (10 $\mu m$) leads to a distribution of particles positions 10 $\mu m$ narrower than the channel width (40 $\mu m$). 

In conclusion, we have demonstrated experimentally that the localization of flowing optical absorbers could be used to obtain photoacoustic images of flow structures with a resolution beyond the acoustic diffraction limit. The limitations of our proof-of-principle study are  the same as that suffered by all the initial proof-of-principle studies which exploited localization to perform imaging beyond the diffraction limit. The main limitation concerns the fact that the localization of individual beads was possible only because we considered a flow of diluted particles. In most practical situation such as imaging the blood vascularization, it will be necessary to find ways to isolate targeted particles before localization. Exogenous contrast agents may be diluted in a flow of red blood cells,  but their photoacoustic signals would have to be sorted from those of red blood cells (RBC). One possible way, as proposed in~\cite{iskander2017photoacoustic}, could be to use difference in absorption spectra to isolate exogenous contrast agents from RBC. Spatio-temporal filtering such as used for ultrasound localization~\cite{demene2015spatiotemporal,errico2015ultrafast,desailly2016contrast} may also be a promising approach to isolate individual RBC inside a dense flow of RBC. A second major limitation is related to the large amount of localization events required to sample the object to image with a sufficient spatial frequency, which leads to a generally very low temporal resolution of localization-based imaging technique independently of the imaging modalities (optical imaging, ultrasound imaging, photoacoustic imaging). Despite these current limitations, which have been or are still actively addressed in the context of other imaging modalities, localization provides a promising way to push the resolution limits of photoacoustic imaging, as it did in recent years for both  optical and ultrasound imaging.

\begin{figure}[htbp]
\centering
\fbox{\includegraphics[width=\linewidth]{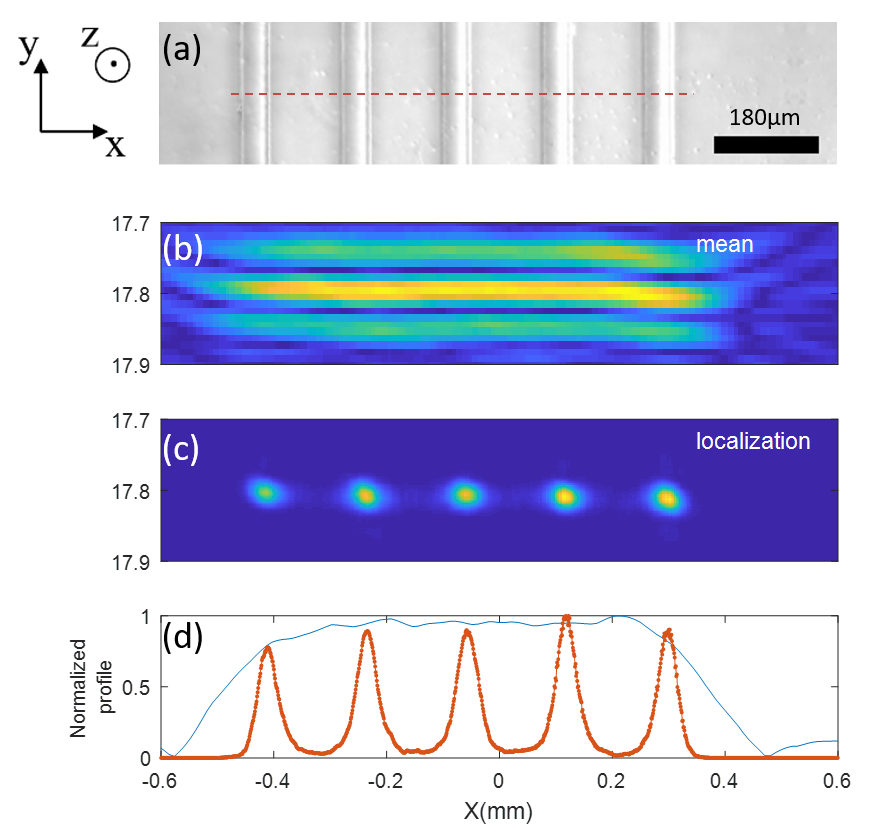}}
\caption{Experimental results. (a) photograph of the sample to image, consisting of 5 parallel microfluidic channels. (b) conventional cross-sectional photoacoustic image of the  5 channels. The resolution of the imaging system is too poor to resolve neighboring channels. (c) Localization image, spatially smoothed with 25 x 25 $\mu m^2$ averaging kernel. (d) Transverse profile across the localization image. The width of each channel as measured by the FWHM ranges from 40 to 45 $\mu m$, for an expected value of $40\ \mu m$.} 
\label{fig:false-color}
\end{figure}

Note: During the finalization of our manuscript, a manuscript posted on the arXiv a few days before our own submission reported similar results obtained with localization-based photoacoustic imaging. The reference list below was therefore updated accordingly just before our submission (see~\cite{dean2017localization}).

\section*{Funding Information}

This project has received funding from the European Research Council (ERC) under the European Union’s Horizon 2020 research and innovation program (grant agreement No 681514-COHERENCE). The authors thank Ori Katz for stimulating discussions.
\bibliography{References}
 

\end{document}